%Paper: hep-ph/9211230
%From: "VSTST2::COMELLI"@BNLDAG.AGS.BNL.GOV
%Date: Mon, 9 Nov 1992 11:12:57 -0500 (EST)

\\
{\Large \bf RADIATIVELY CORRECTED BOUND ON THE } \\[1.2ex] {\Large \bf LIGHT
HIGGS MASS IN A MINIMAL NON } \\[1.2ex] {\Large \bf MINIMAL SUPERSYMMETRIC
STANDARD MODEL } \\[10ex]
\\
{\large D.COMELLI }                                           \\[1ex]
{\it \ Dept. of Theorical Physics,University of Trieste }  \\[4ex]
{\it and INFN sez.di Trieste }                        \\[9ex]
\\
{\large \bf Abstract}
\end{center}
\begin{quotation}
 The neutral Higgs sector of the Minimal non Minimal
Supersymmetric Standard Model is considered. By effective potential
and R.G.E. supported method; an upper bound of the lightest
Higgs is analysed. From the request of perturbativity of the coupling in the
superpotential, adding the leading stop top contributions, the absolute
bound of $\sim 130$ GeV for $90\,GeV\,< m_t<\,180\,GeV$ and
$M_{\tilde{t}} \simeq 1000 $ GeV is derived.
The interesting dependence on $m_t$ for $\tan{\beta} \rightarrow 1$ is
discussed .
\end{quotation}
\end{titlepage}
%%%%% TITLE PAGE
%------------------------------------------------------------------------
The Minimal Supersymmetric Standard Model [1] (M.S.S.M.) is a simple
and attractive phenomenological alternative to the Minimal Standard
Model (M.S.M.), where the disturbing problem of the divergent radiative
 corrections to the Higgs mass is solved in a elegant way, with a minimal
 content of particles.
It has the nice feature that, although three neutral (and two charged)
Higgs bosons appear in the spectrum, an upper bound
exists for the mass $M_h$ of one (the "light") scalar.
At tree level, this reads
\begin {equation}
M_{h}\leq M_Z
\end {equation}
and one expects that radiative corrections will not modify the bound
drastically.
In fact, the existing calculations at one loop [2] show that in the MSSM
there will always be a relatively light scalar of a mass of order O($v$),
where $v=\sqrt{v_1^2 + v_2^2} \simeq 174$ GeV ($v_{1,2}$ are the vevs of the
two doblets
$H_1,H_2$ in the model).

The existence of an upper bound of O($v$) on the lightest Higgs
 mass is  a general property of SUSY models [3]. To fix more quantitatively
the precise value depends strongly on the model and the introduction of the
radiative corrections is very important for this purpose.
In particular it is very interesting to study the modifications induced by
minimal
extensions of the MSSM.
With the introduction of one additional Higgs singlet field N
it is possible to form   a superpotential
\begin{equation}
W=\lambda H_1 H_2 N - \frac{k}{3} N^3
\end{equation}
which can generate in a "natural" way, via the vev $x=<N>$, a term
$\sim \mu H_1 H_2$ whith $\mu=\lambda x \simeq O(M_W)$, difficult to
explain in the MSSM [4].
The price that one has to pay is the appearance
of the extra parameters $x$ and k (necessary to avoid the appearance of one
unwanted Goldstone boson)
and of two extra (a scalar and a pseudoscalar) Higgs bosons.
Quite interestingly, in the neutral  scalar sector a bound for the lightest
Higgs
mass can still be obtained.
The derivation of the bound at tree level is relatively simple, and it has been
provided by a
number of authors [5].
It reads:
\begin{equation}
M_{h}^2 \leq (\frac{g_z^2}{2} \cos^2{2 \beta} + \lambda^2 \sin^2{2 \beta}) v^2
\end{equation}
where $M_z^2=g_z^2/2 \; v^2$ and $\tan{\beta}=v_2/v_1$,
so that, in order to give a quantitative number, some information on
$\lambda$ is now necessary.
In general, one derives bounds on $\lambda$ from
renormalization group equation (RGE) (and request of a perturbative treatment)
arguments.
Once an upper bound on $\lambda$ is assumed, limits on $M_h$ are
consequently derived, that are intuitively still of O($v$) i.e. equal to
$M_z$ or a few more GeV.

One can also evaluate radiative corrections to the bound in the model.
Naively, one would expect the appearance of a "leading" correction
$ \sim m_t^4/M_z^2 \log{M_{\tilde{t}}^2/m_t^2}$ ($m_t ,M_{\tilde{t}}$ are
the mass of the top and of the stop), like in the MSSM case,
leading to a rising of the bound when $m_t$ gets larger.
On the contrary as shown in  a recent paper by J.R.Espinosa and M.Quiros
[6], the plot of the upper bound of $M_h$ versus $m_t$ is strongly dependent
on the value of $\tan{\beta}$ and in particular for $\tan{\beta} \rightarrow 1$
it shows  a pronounced decrease with $m_t$, contrary to the intuitive
expectation.
This seems very interesting from the phenomenological point of view.
Therefore, I have studied in this paper this model with the method of the
effective
potential [7]  to compare the results with those of ref. [6] based on a pure
RGE approach.

Starting from the Higgs superpotential (2), the tree level scalar potential is:
\begin{equation}
V(H_1,H_2,N)= V_F + V_D + V_{SB}
\end{equation}
\begin {equation}
V_{F} = \lambda^2 [ |N|^2 ( |H_1|^2 + |H_2|^2 ) + |H_1H_2|^2 ] +
k^2 |N|^4 - ( \lambda k H_1 H_2 N^{*2} + h.c.)
\end{equation}
\begin{equation}
V_D= \frac{ g_z^2}{8} ( |H_1|^2 - |H_2|^2 )^2 + \frac{g^2}{2} |H_1^{+} H_2|^2
 \end{equation}
and we have assumed the usual expression of the soft supersymmetry
breaking term:
\begin{equation}
V_{SB}= m_1^2 |H_1|^2 + m_2^2 |H_2|^2 + m_N^2 |N|^2 - (A_{ \lambda}
 \lambda N H_1 H_2 + h.c.)- ( \frac{k}{3} A_k N^3 + h.c.)
\end{equation}
where, in absence of explicit CP violation, $\lambda$ and k are real.
Minimizing the potential introduces the real vevs
$v_1,v_2$ and $x$.
The extra vev $x$ is not strongly constrained from  experiment ( differently
from the vev associated to the breaking of one extra U(1) group [8]);
the most interesting bound comes from the lightest chargino mass:
\begin{equation}
M_{chargino} \simeq \lambda x > 45 GeV
\end{equation}
that  implies values of $x\sim$ O($v$).
However, when drawing  the plot of $M_h$ versus $x$ one finds an
 increase with $x$
wich is saturated for $x\gg v$, see fig.[1], so that it is convenient,
to perform a
calculation of the maximum, to work in the  limit $x \rightarrow \infty$ .

The tree level matrix elements $m_{ij}^2$ of the neutral scalar fields ($Re
H_1,
Re H_2, Re N$ ) after  imposing  the minimum conditions are:
\begin {equation}
 m_{11}^2= \hat{m}_{11}^2+\bar{m}_{11}^2= \lambda x \tan{\beta}
( k x + A_{\lambda}  ) + \frac{g_z^2 v_1^2}{2}
\end {equation}
\begin {equation}
m_{22}^2= \hat{m}_{22}^2+\bar{m}_{22}^2=\lambda x \cot{\beta}
( k x + A_{\lambda} ) +  \frac{g_z^2 v_2^2}{2}
\end {equation}
\begin {equation}
m_{12}^2= \hat{m}_{12}^2+\bar{m}_{12}^2= - \lambda x ( k x +A_{\lambda} )+2 v_1
v_2 ( \lambda^2 -
\frac{g_z^2}{4} )
\end {equation}
\begin {equation}
m_{13}^2=\hat{m}_{13}^2+\bar{m}_{13}^2=2 \lambda^2 x v_1 -2 \lambda k v_2 x
-\lambda A_{\lambda} v_2
\end{equation}
\begin{equation}
m_{23}^2=\hat{m}_{23}^2+\bar{m}_{23}^2=2 \lambda^2 x v_2
  - 2 \lambda k v_1 x - \lambda A_{\lambda} v_1
\end{equation}
\begin{equation}
m_{33}^2= \hat{m}_{33}^2+\bar{m}_{33}^2=4 k^2 x^2 - k A_k x +
 \lambda A_{\lambda} \frac{v_1 v_2}{x}
\end{equation}
where the terms $\hat{m}^2$ are of order O($x$) and $\bar{m}^2$ of order
O($v$).
In the limit $x \gg v$ and allowing the soft term A to be  at most of
 order O($x$) the lightest Higgs mass results up to terms O($v/x$):
\begin{equation}
M_{h}^2 (x \rightarrow\infty) = \frac{1}{v^2} [\, \bar{m}_{11}^2 v_1^2+
\bar{m}_{22}^2 v_2^2 +2 \bar{m}_{12}^2 v_1 v_2 \,]-
\frac{[ v_2 \hat{m}_{23}^2 +  v_1 \hat{m}_{13}^2]^2}{v^2 \hat{m}_{33}^2}=
\end {equation}
$$
M_Z^2 \cos^2{2 \beta} + \lambda^2 v^2 \sin^2{2\beta} -
\frac{4}{4-a} \, \frac{\lambda^4}{k^2 } [1- \frac{k}{\lambda} \sin{2 \beta}-
\frac{b}{2} \sin{2 \beta}]^2
$$
where $A_k=k a x$ and $A_{\lambda}=\lambda b x$ ( to have a positive
spectrum $b\leq2$, $a\leq3$, see ref.[9]).
This expression has a maximum for:
\begin{equation}
\hat{z} \equiv \frac{k}{\lambda}=\frac{1}{\sin{2 \beta}} -\frac{b}{2}
\end{equation}
This means that at tree level the bound on $M_h$ becomes:
\begin{equation}
M_h^2 \leq M_z^2 \;\;for\;\; \lambda^2 < \frac{g_z^2}{2} \;\;and\;\;
\tan{\beta} \gg 1
\end{equation}
\begin{equation}
M_h^2 \leq \lambda^2 v^2 \;\;for\;\; \lambda^2 > \frac{g_z^2}{2} \;\;and \;\;
\tan{\beta} \rightarrow 1
\end{equation}
One sees here a major difference with the MSSM and a possible way
out in the case of a not uncovered very light Higgs.
The bound (17) is the same result as that of the MSSM whereas eq. (18)
requires a careful evaluation of the coupling $\lambda$ (
it's interesting to note that in the case in which the extra Higgs is
associated to an extra gauge group
U(1) the dependence on $\lambda$ disappears
automatically in the maximization of
$M_h$, see ref.[9]).

Consider now the introduction of the leading radiative corrections,
coming from the top stop sector. For the upper bound
 only the corrections to the elements $(11),(12),(22)$ are required becouse
the maximum at tree level contains only the first term on the right side of
eq.(15).
The  field dependent masses which one must introduce in the effective potential
are:
\begin{equation}
M_{\tilde{t}_{R,L}}^2=m^2+(-)\Delta^2
\end{equation}
\begin{equation}
m_t^2=h_t^2 v_2^2
\end{equation}
\begin{equation}
m^2\equiv m_{soft}^2 + m_t^2 + \frac{g_z^2}{8} (v_1^2 - v_2^2)
\; ; \; \Delta^2 \equiv h_t (\lambda x v_1 - A_t v_2)
\end{equation}
with $h_t$ the top Yukawa coupling and  $m_{soft}^2,A_t$ the soft mass
SUSY breaking terms.
The radiative corrections to each element are:
\begin{equation}
\delta m_{11}^2 = \frac{3}{32 \pi^2}[(\frac{g_z^4}{16} v_1^2 +
\lambda h_t^2 A_t x \tan{\beta})
Z + g_z^2 \lambda h_t v_1 x \frac{\Delta^2}{m^2}]
\end{equation}
\begin{equation}
\delta m_{12}^2= \frac{3}{32 \pi^2}[((2 h_t^2- \frac{g_z^2}{4})\frac{g_z^2}{4}
v_1 v_2 - \lambda h_t^2 A_t x )
Z + 2 \frac{\Delta^2}{m^2} ( \lambda h_t v_2 x ( 2 h_t^2 - \frac{g_z^2}{4}) -
\frac{g_z^2}{4} h_t A_t v_1)]
\end{equation}
\begin{equation}
\delta m_{22}^2 = \frac{3}{32 \pi^2}[(\frac{g_z^4}{16} v_2^2
- g_z^2 h_t^2 v_2^2 +
\lambda h_t^2 A_t x \cot{\beta})Z + 4 h_t^4 v_2^2 (
\log{ \frac{  M_{\tilde{t}_R}^2  }{m_t^2} } +
\end{equation}
\begin{displaymath}
\log{ \frac{  M_{\tilde{t}_L}^2 }{m_t^2} })- 4 \frac{\Delta^2}{m^2}
 h_t A_t v_2
(2 h^2_t - \frac{g_z^2}{4})]
\end{displaymath}
here I have used the simplification $\log{(M_{\tilde{t}_R} / M_{\tilde{t}_L})}
\simeq 2 \Delta^2 / m^2$ and the position
$Z=\log{(M_{\tilde{t}_R}^2 / M_z^2)} + \log{(M_{\tilde{t}_L}^2 / M_z^2)}$.

The radiatively corrected mass is then
\begin{equation}
 M_h^2=\frac{1}{v^2}[\bar{m}_{11}^2v_1^2 + \bar{m}_{22}^2 v_2^2 +
2 \bar{m}_{12}^2 v_1 v_2 + \delta m_{11}^2 v_1^2 +
\delta m_{22} v_2^2 + 2 v_1 v_2 \delta m_{12}^2]
\end{equation}
In agreement with the screening equation of ref. [3],valid in
the large $x$ limit, the radiative corrections of
order O($\alpha x$) cancel and  the corrected upper bound becomes :
\begin{equation}
M_h^2=M_z^2\cos^2{2 \beta}+ \lambda^2 v^2 \sin^2{2\beta} +
\frac{3}{32 \pi^2} v^2 [(\frac{g_z^4}{16}\cos^2{2 \beta} + g_z^2 h_t^2
\sin^2{\beta} \cos{2 \beta})Z +
\end{equation}
$$
\frac{\Delta^4}{v^2 m^2}(8 h_t^2
\sin^2{\beta} + g_z^2 \cos{2 \beta}) + 4 \frac{m_t^4}{v^4}
(\log{\frac{ M_{\tilde{t}_R}^2}{m_t^2}}+
\log{\frac{ M_{\tilde{t}_L}^2}{m_t^2}})]
$$
A precise evaluation of the maximum of $M_h$ goes through an analysis
of the possible value of $\lambda$.
It becomes necessary, for such purpose, to perform a careful analysis
of the RGE [5] and of the parameter space involved in the numerical evaluation
of the mass of
the light Higgs.
The R.G.E. promote the configurations in which the free parameters of the
superpotential (2), $k$ and $\lambda$, are in the ratio
\begin {equation}
\frac{\lambda^2}{k^2}=2
\end{equation}
as shown by the paper of P.Binetruy and C.A.Savoy [5]; at the same time
a contourplot of $M_h$ (radiatively corrected) on the  ($k,\lambda$) plane
shows that a long this direction the maximum is reached most quickly,
see fig.[3].
So imposing this constraint, the RGE reduce to:
\begin {equation}
\frac{d h_t}{d t} \simeq \frac{h_t}{8 \pi} (3 h_t^2 - \frac{8}{3} g_s^2)
\end{equation}
\begin{equation}
\frac{d \lambda}{d t} \simeq \frac{\lambda}{8 \pi} ( \frac{5}{2} \lambda^2 +
\frac{3}{2} h_t^2 - \frac{3}{2} g^2 - \frac{1}{2} g^{'2})
\end{equation}
($ t  =\log{ \frac{\mu}{M_z} }$) , with $g_s$, $g$ and $g^{'}$ the couplings
of the gauge groups SU(3), SU(2), U(1).

Fig.[5] gives the result of the numerical analysis in which I have imposed the
constraint of perturbativity on $\lambda$ ($\frac{\lambda^2(\Lambda)}
{4 \pi} \leq 1$ with $\Lambda \sim 3 \, 10^{16}$ GeV) and shown  how
$\lambda[M_z]$ strongly depends on the values of $h_t[M_z]$ setting a
severe bound on it.

In this way the dependence of $M_h$ on $m_t$ is double; an explicit one,
coming from the effective potential, and an implicit one in the running of
the elements at tree level.
Whereas the first one tends to increase the upper bound for growing $m_t$, the
second one, on the contrary, tends to decrease it.
The relative weight of this two contributions depends on the
values of $\tan{\beta}$ and the first is larger for $\tan{\beta}\gg1$,
(the parameters space of eq. (17)), the second for $\tan{\beta} \rightarrow 1$.
In such a way there are two different pattern of $M_h$ on $m_t$, one increasing
for
$\tan{\beta}\gg1$, very similar to the values of the MSSM bound, another one
strongly decreasing for $\tan{\beta} \rightarrow 1$.
At intermediate regions this two effects compensate, from which the relative
flatness of the intermediate curves of fig.[5] emerges.
Making the envelope of all the curves, $M_h$ results always smaller than 130
GeV
for $90 GeV <m_t < 180 GeV$,
a result that shows a difference of $\sim 8 $ percent  from that obtained with
  the
method of ref. [2]confirming the relative stability ot the outcome with
respect to various reasonable theoretical imputs.

Acknowledgements

I like to thank C.Verzegnassi, who inspired this work and followed each
stage of its preparation. I thank also P.Furlani for the assistance
during this work.

\vfill
\newpage

REFERENCES

[1] See e.g. for a discussion of this point L.E.Ib\'{a}\~{n}ez ,
$CERN-TH.5982/91 $ and references therein.

[2] Y.Okada, M.Yamaguchi, T.Yanagida, Prog.Theor.Phys. 85 (1991) 1

J.Ellis, G.Ridolfi, f.Zwirner, Phys.Lett.B 257 (1991) 83

R.Barbieri, M.Frigeni, F.Caravaglios, Phys.Lett. B 258 (1991) 167

[3] D.Comelli, C.Verzegnassi, preprint DESY 92-109

[4]E.Kim, H.P.Nilles, Phys.Lett. 138 B (1984) 150
B
L.Hall, J.Lykken, S.Weinberg, Phys.Rev. D 27 (1983) 2359

[5] M.Drees, Int.J. of Mod.Phys. A, Vol 4, No.14 (1989) 3635

P.Bin\'{e}truy, C.A.Savoy, Phys. Lett. B 277 (1992) 453

U.Ellwanger preprint HD-THEP-91-21

[6] J.R.Espinosa, M.Quiros, preprint IEM-FT-60/92

[7]S.Coleman, E.Weinberg, Phys. Rev. D 7 (1973) 1888

G.Gamberini, G.Ridolfi, F.Zwirner, Nucl.Phys. B 331 (1990) 331

[8] J.Ellis, D.V.Nanopoulos, S.T. Pectcov, F.Zwirner Nucl.Phys. B 283 (1987) 93

 H.E.Haber, M.Sher, Phys.Rev. D 35 (1987) 2206

 V. Barger, K.Whisnant, Int. J. of Mod. Phys. A, Vol.3, No.8 (1988) 1907

[9] D.Comelli, C.Verzegnassi preprint  DESY 92-087

\vfill
\newpage

FIGURE CAPTIONS

Fig.[1]   Plot of $M_h$ versus $x$ from a numerical diagonalization of the 3x3
neutral mass matrix, radiatively corrected, for different values of
$\lambda$ [ k=0.5; $A_k = A_{\lambda}=100$ GeV, see ref.[5]; m=1000 GeV;
$\Delta$=400 GeV; $m_t$=150 GeV; $\tan{\beta}=20$ ].

Fig.[2]  The same as before but with k=0.6; $\tan{\beta}$=1

Fig.[3]   Contour Plot of $M_h$ in the plane (k,$\lambda$);
    $x=1000$ GeV;     $A_k=A_{\lambda}=$100 GeV; m=1000 GeV; $\Delta$=400 GeV;
$m_t$=150 GeV; $\tan{\beta}$= 1.
The dotted straight line represents the fixed ratio $\lambda^2 / k^2 =2$.

Fig.[4]   The same as before but with $\tan{\beta}$=20

Fig.[4]   Plot of $\lambda^2 [M_z]$ versus $h_t[M_z]$ from the request of
perturbativity at the scale $\Lambda \sim 3\,10^{16}$ GeV and with the
constraint $\lambda^2 / k^2$=2.

Fig.[6]   Plot of the upper bound (eq.(26)) versus $m_t$ with $\lambda$
running (fig. [4]) for different values of $\tan{\beta}$.

\vfill
\end{document}